# Directed percolation in nuclear safety

V. V. Ryazanov

Institute for Nuclear Research, pr. Nauki, 47 Kiev, Ukraine, e-mail: vryazan19@gmail.com;
ORCID ID: 0000-0002-5308-3212

Neutron behavior in a nuclear reactor is described using a directed percolation model. The preferred direction is created by time-oriented neutron generations. Using the time required to reach a dangerous neutron flux limit or reactor power limit as an example, it is shown that, in certain situations, the proposed approach can detect events hazardous to reactor safety that are undetectable by traditional reactor safety systems. The capabilities and limitations of the proposed approach are outlined.

## 1. Introduction

A number of works, such as Refs. [1-8], are devoted to the study of stochastic phenomena in the neutron system of a nuclear reactor, including neutron noise. These include fundamental works on the stochastic nature of neutron fields and experiments confirming this nature. In [6], direct measurements of noise in the core were conducted. The spectrum slope in the low-frequency region corresponds to the parameter $a$ used below in formula (9) for developed nucleate boiling. This article assumes the probabilistic nature of processes in a nuclear reactor and uses stochastic methods for their study, as was done in Ref. [9].

The classical theory of neutron noise, developed in Refs. [1-8], convincingly demonstrates that fluctuations in the parameters of the environment in power reactors (primarily due to boiling) exceed the intrinsic noise of the fission process by orders of magnitude. Experimental comparison of models based on inverse Kolmogorov equations with spectral analysis data confirmed the presence of long-period correlations. Our study expands this approach, moving from a local description of noise to a global analysis of temporary safety reserves at the critical point of directed percolation.

However, neutron noise theory does not consider a situation that is perhaps the most dangerous for nuclear reactors: distributions with "heavy tails," power-law distributions with negative exponents less than 2. These types of distributions are characteristic of directed percolation processes (Refs. [10-14]).

Directed percolation (DP) differs from conventional isotropic percolation in that it has a distinct percolation direction. There is a strictly defined direction (vector), most often interpreted as a time axis. The resulting structures appear differently. Anisotropic symmetry (different along and across the axis of motion) arises. DP and isotropic percolation belong to different universality classes. Mathematically, this is described by different critical exponents. DP adds a causality factor (the only possible path is "forward"). In DP, a neutron in generation $n$ gives birth to descendants in generation $n$+1, creating an irreversible chain of cause-and-effect relationships. At the critical point $k_{eff}$=1, the system undergoes a second-order phase transition, where the correlation scale theoretically becomes infinite. In Ref. [15], it is shown how fractional derivative equations arise due to DP and super-diffusion. Non-classical values of the DP indices are calculated in Ref. [10]. In Ref. [15], the transition from the mean-field approach to a realistic three-dimensional geometry with a critical size $d_c$=4 is described. It is demonstrated that fractional dynamics is an inevitable consequence of reactor operation near criticality, confirming the need for fractional indices specific to the universal class of DP.

This article is devoted to directed percolation and reactor safety. However, many other issues are related to the problems raised in the article. For example, it is proposed to use the apparatus of boundary functionals of random processes Ref. [9] to consider reactor safety issues. Power-law distributions and stable distributions Refs. [16-17], whose physical aspects are self-similarity and fractality Refs. [18-21], represent a transition to directed percolation Refs. [10-14]. This is where reactor safety comes into play. Under nominal water-moderated power reactor (WWER) reactor conditions, the high density of the moderator (water) effectively suppresses the "heavy tails" of the distributions (for neutron ranges or for the number of effective descendants of a single particle in all generations), limiting the discrepancy between the DP model and classical diffusion to 1–2% Ref. [15]. Power-law distributions are possible in other situations, for example, for other reactor types or for certain critical assemblies. Specifically, for WWER reactors, this is relevant for emergency situations involving strong core heterogeneity. Directed percolation and Levy statistics are tools for analyzing catastrophes (core destruction) or for designing fundamentally different systems (molten salt reactors (MSRs), high-temperature gas-cooled reactors (HTGRs), and pulverized fuel reactors).

In a standard WWER reactor, the distribution of neutron ranges and the number of secondary neutrons have a finite variance. According to the central limit theorem, this leads to a normal distribution and the standard diffusion equation.

The transition to power-law distributions and directed percolation occurs in the following cases: a). Stochastic (granular) media: if the core consists of randomly distributed fuel microspheres (e.g., in high-temperature gas-cooled reactors) or contains random cavities. b). Strong density fluctuations (boiling water reactors): in zones with intense steam formation, steam bubbles can form "fractal labyrinths" through which neutrons pass without collisions over anomalously large distances. c). High-burnup fast neutron reactors: where localized "oases" of fissile material are formed within the absorber mass. The use of directed percolation and Lévy distributions is crucial if the biological shield contains microcracks, whose distribution may follow a power law. In this case, the radiation does not propagate exponentially (Bouguer's law), but rather much more slowly, "leaking" through the medium. Another situation with DP is related to neutron noise. Near the critical point, reactor power fluctuations become abnormally large. If the distribution of descendants P(k) has a "heavy tail," standard monitoring systems may fail to respond to a localized power spike (a "neutron burst"), since the average value is no longer representative.

An important question concerns the meaning of the parameters s and k from equations (8), (13), (17). When s=0, k(s)=0, phase transitions with divergent characteristics occur. But in Refs. [22-23] it is shown that in expressions (8), (13) one should not assume that s=0, as is done when defining moments in the theory of random processes, since s=0 only at equilibrium (just as $\beta=1/kT=0$ only at $T\to\propto$, where k is the Boltzmann constant, T is the temperature; β and energy E are conjugate quantities. Similarly, s and $T_{FPT}$ are conjugate quantities in (12); the kinetic equations for the distributions of E and for the distributions of the first-passage time to reach the level (FPT) of random value $T_{FPT}$ are also conjugate), s~1/t, t is the time when s=0, $t\to\propto$, but the moment t must belong to the FPT interval and be finite. Time is inseparable from space. In the relations given below, we pass from boundedness in time to boundedness in space.

The article is organized as follows. Section 2 discusses the relationship between directed percolation and nuclear safety. Section 3 provides mathematical definitions and justifies the use of a power-law distribution. Section 4 examines the effect of finite reactor dimensions. Section 5 examines the relationships between the FPT functional, the Lundberg equation, and the canonical form of the characteristic functions of stable distributions. Section 6 examines neutron processes in the reactor and the critical indices of the DP. Section 7 calculates the dependence of FPT on the critical exponent. Section 8 concludes. Appendix A contains the derivation of expressions for FPT and approximations for critical points.

## 2. Directed percolation and nuclear safety

During startup and at minimum power levels, the stochastic nature of the neutron spectral density is most pronounced. Low neutron density and the absence of pronounced feedback (the Doppler effect) make the system vulnerable to local "neutron bursts" generated by Lévy statistics. To ensure the stability of control systems, it is necessary to consider not only average power values but also the highest statistical moments of neutron noise. The detection of power-law dependences in the spectral power density can serve as an early indicator of the system's transition to an unstable percolation regime.

The WWER system is too "dense" and "regular" for power laws. DP and Lévy statistics are tools for analyzing accidents (core collapse) or for designing fundamentally different systems. In Ref. [15], it is shown that the critical point of a nuclear reactor corresponds to a DP phase transition. Subsequent studies develop the ideas of Ref. [15] through the analysis of "neutron clustering" of Ref. [24], the use of stochastic "predator-prey" models to account for feedback, and the application of DP physics to Monte Carlo convergence problems. Ref. [15] shows that, in steady-state conditions, classical calculations based on the DP yield only a 1–2% difference compared to fractional models. However, this approach opens up fundamentally new aspects of neutron behavior in reactors. The risks associated with underestimating local criticality are explored in papers on "neutron clustering" (Ref. [24]), an analysis of how neutrons "clump" together in space, creating local power peaks. Unexpected criticality can arise from density fluctuations. Ref. [24] discusses in detail why "neutron clusters" arise in Monte Carlo simulations, how the central limit theorem is violated, and how spatial correlations similar to percolation ones appear. In the DP model, "neutron bursts" occur instead of a smooth profile. During neutron clustering (the "hub effect" in DP), the local fission density at one point in the fuel element can exceed the calculated value by several times for milliseconds. The central limit theorem is valid for the Gaussian distribution. Other stable distributions also belong to the class of infinitely divisible distributions, which, like the Gaussian distribution, are limit laws to which the sums of a large number of distributions converge (Ref. [16]). In Ref. [25] experimental evidence is given for the existence of "spotted" nuclear chain reactions, showing that the neutron field at low powers consists of isolated clusters rather than a uniform background. Ref. [25] highlights the physical parameters of clustering, demonstrates the inapplicability of standard diffusion models to describe the real risk, and confirms the connection of these phenomena with DP using experimental data. There are a number of mathematical papers on DP and its connection with the Levy-Khinchin canonical form (Refs. [10-14]). Ref. [25] is a fundamental point of contact between the abstract theory of DP and a real experiment in nuclear physics. It investigated the transition from "gas" to "clusters" and proved that in a critical reactor neutrons are distributed not as a rarefied gas (diffusion), but as spotty structures. In [25] it is shown that due to the branching (fission) process, neutrons have common "ancestors", which causes them to stick together in groups.

This is a direct manifestation of the DP theory in space-time: fissions chains form tree-like structures that become fractal near $k_{eff}$=1. Experimental confirmation has been obtained (Experiments at LANL). This is a rare case of testing the DP theory on a real critical assembly (Planet, Los Alamos). The authors measured correlations between fissions and found that the spatial structure of the flux fluctuates anomalously. The statistics of these fluctuations are not described by a Poisson (classical) distribution. They exhibit "heavy tails," which directly indicates Lévy statistics and effects near $a$=2 in the power-law distribution of $k^{-a}$ (9). Although the article is physical in nature, it relies on the powerful mathematical apparatus of branching Brownian motions. At the microscopic level, a lattice or graph (the DP model) is considered, where particles fission and die. A mathematical connection with the canonical Levy-Khinchin form is traced. The work effectively demonstrates that the probability density of neutrons in a cluster is described by stable power laws. The sum of many fission events in a "spotted" medium leads to a distribution whose characteristic function has the form given by the Levy-Khinchin formula (with a fractional exponent α). This explains why "jumps" in power occur: a cluster can instantaneously "grow" over a large distance. It should be noted that power laws for the critical region were obtained in Ref. [26]. For the distribution function, the probability P(k) is written for the number of descendants or jump lengths with a power-law tail of the form $k^{-a}$ (9). The characteristic function $\Phi(t)=E(e^{itX})$ is calculated. For power-law tails, it has a non-analytic form at zero: $\ln\Phi(t)\sim-|t|^{\alpha}$. This structure $\ln\Phi(t)$ exactly corresponds to the Lévy-Khinchin representation for stable distributions. When passing to the continuous limit (differential equation) at the macroscopic level, this term $|t|^{\alpha}$ turns into a fractional derivative with respect to space or time. Ref. [25] is the "physical facade" showing that nature is indeed structured this way. Refs. [10] and [14] are the "mathematical foundation" proving that this is an inevitable consequence of the violation of the central limit theorem in branching processes and convergence to other stable distributions. Let us summarize the results of modern research in applied reactor physics. Ref. [25], combining the DP theory and real experiments on critical assemblies, demonstrates the existence of neutron clustering ("spotting"), which is a physical manifestation of the DP in space-time. Ref. [15] applies dynamic DP indices to the calculation of reactor systems. The authors use fractional derivatives to describe the evolution of the neutron field, confirming that at critical points, classical diffusion requires a correction for DP effects. The transition to the description of a reactor through the prism of DP allows for a deeper interpretation of the physics of fluctuations and "black swans" in nuclear power, providing a scientific basis for the safety analysis of promising and highly heterogeneous cores.

During normal operation, the probability of the local neutron distribution becoming a power law ($a$=2) is virtually zero due to the homogeneity of the coolant. The probability of conditions for directed percolation in the melt is estimated at approximately $10^{-5}$ to $10^{-6}$ per reactor per year (this is the probability of the most severe accident). However, it's not the probability of occurrence that's important, but the error in the calculations. If an accident occurs, and we consider the melt to be a normal diffusion code, we obtain a "safe" configuration. However, if we apply the $a$=2 model, it turns out that due to "hubs" (fuel accumulations) and "shoots" (Lévy flights), the system can reach prompt neutron emission.

## 3. Mathematical description and power laws

As noted in the introduction, stable power laws have not been investigated in reactor noise. Let us define the quantities that will be used below. We define First-passage time (FPT) as the time it takes for $\tau^+(x)$ to reach a positive level x>0 by the process ξ(t) Ref. [9]. Below, only positive levels will be considered, and the index + will not be used;

$\tau^+(x) = \inf\{t : \xi(t) > x\},\ \ x > 0$ is the moment of the first exit for the level $x$>0. (1)

In our case, a more specific definition is used below:

$$T_{FPT}=\inf\{t>0:\ \Phi(t)\geq\Phi_{crit}\}, \tag{2}$$

where Φ(t) is the neutron flux or reactor power, considered as a random process, and $\Phi_{crit}$ is some critical value of this quantity. Definition (1) can also be applied to other reactor characteristics, such as the service life of fuel elements, etc.

A process $\left\{\xi\left(t\right),\ 0\leq t\leq T\right\}$ is called a process with independent increments if for an arbitrary $n$≥1 and for all $\{t_k\}_{k=\overline{0,n}}$, $0 \leq t_0 < t_1 < ... < t_n \leq T$, the values $\xi(t_0)$ and increments $\xi(t_0)$, $\xi(t_1)$ - $\xi(t_0)$,…, $\xi(t_n) - \xi(t_{n-1})$ are independent random variables.

A process with independent increments is called homogeneous if the distribution of increments $\xi(t) - \xi(s)$ depends only on $t$-$s$ when $s \leq t$. If $\xi(t)$ is a homogeneous process with independent increments, then we will assume that $0 \leq t < \infty$, and it is often assumed that $\xi(0) = 0$.

Let $X$ be a random variable with cumulative distribution function $F_X$. The moment generating function ($MGF$) of $X$ (or $F_X$), denoted by $M_X(t)$, is:

$$M_X(t)=\boldsymbol{E}[e^{tX}] \tag{3}$$

provided this expectation exists for $t$ in some open neighborhood of 0. That is, there is an $h$>0 such that for all $t$ in $-h<t<h$, $\boldsymbol{E}[e^{tX}]$ exists. If the expectation does not exist in an open neighborhood of 0, we say that the moment generating function does not exist.

If $X$ is a continuous random variable, the moment-generating function's (MGF) definition expands (by the law of the unconscious statistician) to:

$$M_X(t) = \boldsymbol{E}[e^{tX}] = \int_{-\infty}^{\infty} e^{tx} f_X(x)dx\,, \tag{4}$$

where $f_X(x)$ is probability density function.

If $\xi(t)$ is a homogeneous stochastically continuous process with independent increments in $R^m$, then its characteristic function takes the form (a Lévy-Khinchin type representation):

$$\boldsymbol{E}e^{i(z,\ \xi(t))} = \exp\{t[\,i(z,a) - \frac{1}{2}(Bz,z) + \int_{|x|\leq 1}(e^{i(z,x)} - 1 - (z,\ x))\Pi(dx) + \int_{|x|>1}(e^{i(z,x)} - 1)\Pi(dx)]\}\,, \tag{5}$$

where $a \in R^m$; $B$ is a symmetric non-negative operator in $R^m$; $\Pi$ is a measure in $R^m$, for which $\int \frac{(x,x)}{1+(x,x)} \Pi(dx) < \infty$, and $\Pi(\{0\}) = 0$.

The characteristic function (CF) of a random variable $\xi$ with a distribution function $F(x) = P\{\xi < x\}$ is referred to as a complex-valued function:

$$\varphi(t) = \boldsymbol{E} e^{it\xi(t)} = \int_{-\infty}^{\infty} e^{itx} dF(x). \tag{6}$$

The characteristic function of a homogeneous process $\xi(t),\ t \geq 0$ is determined in the theory of random processes (for $\xi(0) = 0$) (Ref. [17]) by the relation:

$$\boldsymbol{E} e^{i\alpha\xi(t)} := \int_{-\infty}^{\infty} e^{i\alpha x} dF(x) = e^{t\Psi(\alpha)},\ t \geq 0, \tag{7}$$

where $F(x) = P(\xi < x)$ is the distribution function of a random process $\xi(t),\ t \geq 0$, and the function $\psi(\alpha)$ represents the scaled cumulant generating function (*SCGF*) of the process $\xi(t),\ t \geq 0$. The *SCGF* do not depend on *t* and characterize all finite-dimensional distributions of the process.

If for a process $\xi(t),\ t \geq 0$, the function $\psi(\alpha)$ at $i\alpha = r$ is equal to *s*, then we obtain the equation:

$$\Psi(\alpha)\big|_{i\alpha=r} := k(r) = s, \quad \pm \operatorname{Re} r \geq 0. \tag{8}$$

This equation, in risk theory, is referred to as the fundamental Lundberg equation.

In our model, the number of offspring generated by each node (neutron) is treated as a random variable, denoted k, that follows a given probability distribution P(k). We assume that in some situations, the distribution P(k) follows a stable power law:

$$P(k) \sim k^{-a}. \tag{9}$$

Such a dependence was obtained, for example, in Refs. [27, 28]. In Refs. [15, 24, 25] it was shown that the branching random walk of neutrons in the presence of feedback mechanisms (for example, the Doppler effect) can belong to the universality class of directed percolation. In these models, the critical decay or growth of clusters is described by power laws. The article [15] analyzes the distribution of the total number of descendants of a single neutron (the cluster size in space-time). In the critical state $k_{eff} \approx 1$, this distribution takes on the power-law form (9) $P(k) \sim k^{-a}$, where the exponent *a* is determined by the critical indices of percolation theory. The Ref. [15] proved that the critical point of a nuclear reactor corresponds to a phase transition of directed percolation. The authors (including Ref. [15]) consider not the fission of a single nucleus, but the effective number of progenies in a heterogeneous medium. A single neutron can trigger a chain of fissions in a local fuel cluster ("subcritical breeding"). If the medium is fractal or is near the critical point, then the effective number of neutrons emitted from such a "super-node" can have a very wide spread. It is this "effective" quantity that can behave as a power law on the scale of a macroscopic lattice. Ref. [25] proves that in a critical reactor, neutrons are distributed not as a rarefied gas (diffusion), but as patchy structures. The authors show that, due to the branching process (fission), neutrons have common "ancestors", which forces them to stick together in groups. This is a direct manifestation of directed percolation in space-time: fissions chains form tree-like structures, which become fractal near $k_{eff}=1$. A power-law distribution also holds for neutron travel distances.

The power-law distribution is valid for a critical branching process near a bifurcation point. In the context of a nuclear reactor, it describes the probability distribution of the total number of neutrons produced by a single initial neutron chain (the total chain size *n*) when the reactor is exactly at the critical state $k_{eff} = 1$. In general, this is not the distribution of the number of immediate descendants of a single neutron k, which is usually approximated by a Poisson distribution. It is the distribution over the total number of neutrons in the chain, including all generations up to its extinction. Under special conditions (very small, "pulsating" assemblies, highly absorbing environments where spatial fluctuations are critical), deviations can be observed, and the DP model may be a more suitable one. The neutron's lifetime is short, but it manages to diffuse a significant distance before fission. This effectively "washes out" spatial correlations, increasing the effective dimensionality of the system to the critical value $d_c=4$, where the mean-field exponents (including 3/2) apply. If spatial fluctuations in a 3D reactor are taken into account (the real physical case), the exponent $a$ is corrected by the critical DP indices: $a=1+\beta/(\beta+\nu_{II})$, where for 3D directed percolation: $\beta\approx 0.81$ (density index), $\nu_{II}\approx 1.90$) (correlation time index). This yields a value of $a\approx 1.3$. In addition to the critical point, the transition to a power-law regime occurs in such specific cases as the action of feedback effects: when taking into account nonlinear effects (for example, temperature feedback), which transform the system into a self-organizing critical environment; strong spatial heterogeneity, when the structure of the active zone or neutron fluxes acquires fractal properties. In Ref. [15] it is substantiated that the "effective" number of descendants is not simply the number of fissions in all generations, but the total contribution of the branching process, limited by "leakage" or absorption.

The power law arises because, near criticality, the probability of long chains decreases slowly: instead of exponential decay $\exp[-k/\langle k\rangle]$, we get algebraic decay $k^{-a}$. In a reactor, this explains "neutron clustering" Ref. [25]—zones of anomalously high neutron density that appear "out of nowhere" and then disappear, which is impossible within the framework of standard diffusion.

In Refs. [15, 24, 25] it is emphasized that if the distribution of neutron ranges is itself “heavy” (for example, due to voids in the geometry), then the process goes into the Lévy-DP regime, where the exponent $a$ in (9) begins to depend on the Lévy flight parameter $\alpha$.

In Ref. [15] the dependence $N(x)\sim x^{\beta}$ is written for prompt neutrons in the case of DP. In most physical and statistical systems, where a random variable (for example, the number of nodes or the cluster volume N) is scaled in this way, the probability distribution P(N) also obeys a power law. In Ref. [26] showed that in the critical regime the trajectories of the process for the total number of neutrons behave like $t^{h}$, increasing according to a power law. More precisely, the random process behaves as $x(t)\sim t^{h}$, where h is a random variable uniformly distributed over the interval (0,1), i.e. $P\{h<x\}=x$ over the interval (0,1). The "effective number of descendants" k refers to the total number of neutrons generated across all subsequent generations by a single "parent" particle under specific physical conditions. In a conventional WWER reactor, the distribution of these descendants does not follow a power-law pattern. However, a power-law distribution (type (9)) can emerge under conditions characterized by strong correlations and anisotropy, which are hallmarks of directed percolation. This phenomenon typically occurs in the three scenarios noted in the introduction. The power-law tail exponent ($a=\alpha+1$) of the probability density function of the stable distribution for large values of k in (9) decreases according to a power law. Confusion between $a$ and $\alpha$ can arise due to the form of notation of the probability density function of the stable distribution. The stability index ($\alpha$) determines the properties of the distribution and is included in the characteristic function as $\exp(-|k|^{\alpha})$. It determines the scaling properties (the self-similarity parameter).

If the distribution of the number of effective progeny P(k) obeys a power law $k^{-a}$ (9) with $a\approx 2$, the system leaves the Gaussian universality class (normal distribution). In a real fission event, it is physically impossible to emit 100 or 1000 neutrons ($P(\nu)$ instantly vanishes after $\nu=8$ due to the limited binding energy of the nucleus). In a classical reactor (WWER), the "physical" $P(\nu)$ is a narrow Gaussian/Terrell. DP effects with power-law tails are properties not of the nucleus itself, but of the collective behavior of the neutron gas in a complex geometry.

The value $a=2$ is a "turning point" for statistical moments. If $a\leq 2$, then the mathematical expectation (the average number of offspring) $\langle k\rangle$ diverges (tends to infinity for an infinite system size). If $2<a\leq 3$, then the mean is finite, but the second moment $\langle k^{2}\rangle$ diverges. If $a\approx 3$, then both the mean and the variance are finite. When the variance is finite ($a\approx 3$), the classical central limit theorem applies to the sum of such quantities, and the behavior of the system becomes "typical" in many aspects. For $a\approx 3$ (where variance is finite), the random walk through many steps is described by the ordinary diffusion equation (the normal distribution of coordinates). For $a<3$ (including the case $a=2$), Lévy flights occur.

For distributions with "heavy tails," the mean is on the verge of divergence. This makes the tree extremely heterogeneous: "super-nodes" (hubs) with a huge number of connections will appear, completely destroying the familiar statistics characteristic of a normal distribution. If the distribution of the number of effective offspring P(k) obeys a power law (9) with $a = 2$, the system departs from the Gaussian universality class (normal distribution). Where does the "power law distribution" in DP studies come from? The authors (including Refs. [15], [25]) consider not the fission event of a single nucleus, but the effective number of progenies in a heterogeneous medium: a single neutron can initiate a chain of fission events in a local fuel cluster ("subcritical breeding"). If the medium is fractal or is near the critical point, then the effective number of neutrons emitted by such a "super-node" can have a very wide spread. It is precisely this "effective" quantity that can behave as a power law on the scale of a macroscopic lattice. In a classical reactor WWER, the "physical" distribution is a narrow Gaussian/Terrell distribution. The DP effects with power-law tails are properties not of the core itself, but of the collective behavior of the neutron gas in a complex geometry. The "effective number of descendants" is the total number of neutrons produced in all subsequent generations by a single "parent" particle, i.e., the chain length. For the vast majority of operating reactors (including WWER, boiling water reactor (BWR), and pressurized water reactor (PWR)), the distribution of the effective number of descendants is narrow and has a finite variance (exponential decay of tails). A power-law distribution is an "exotic" state that occurs only under certain physical conditions. In a conventional WWER, the distribution is not a power-law. In a conventional reactor, the environment is too cramped for a neutron. In the moderator (water), the neutron constantly collides with hydrogen. Its trajectory is a tangled ball (Brownian motion). Due to the huge number of small random collisions, the overall statistics of the system obey the central limit theorem and tend toward a normal (Gaussian) distribution. Under these conditions, the probability that an individual fission chain will suddenly become gigantic (for example, $10^6$ times larger than the average) decreases exponentially. This is called "Gaussian truncation" of the tails.

When does the distribution become a power law ($a$=2)? The power-law form $P(k)=k^{-a}$ arises under conditions of strong correlation and anisotropy, characteristic of DP. This is possible in the following cases: a) Critical point (critical opalescence). Only at the point $k_{eff}=1$ (or in an infinitesimal vicinity) does the distribution of chain lengths become a power law. This is a universal property of all systems undergoing a phase transition. But as soon as the reactor goes supercritical (runs away), nonlinear feedbacks (the Doppler effect) immediately "cut off" this tail, returning the system to a predictable shape. b) Strong spatial inhomogeneity ("Lévy glass"). If the medium consists of empty channels and dense fuel blocks (e.g., a destroyed zone or specific experimental assemblies). A neutron can fly a huge distance without collisions ("Lévy flight"). In such an environment, "offspring" are born not as a compact cloud, but as a fractal structure. Here, the geometry of the environment imposes a power law on the offspring distribution. c) Low neutron count (low sample statistics). In launch modes, when the nucleus contains only a few hundred neutrons, averaging does not work. Individual "successful" fission chains may dominate. Studies (e.g. Ref. [15]) show that under these conditions the dynamics are more similar to DP than to diffusion. Where to look for power-law tails? In a WWER reactor at 100% power, the progeny distribution is narrow (Gaussian/Terrell), the $a$=2 model is not applicable, and neutron propagation processes are controlled by diffusion. For a WWER reactor during startup, the distribution is transient (tails appear). The $a$=2 model is partially applicable (for noise analysis). For a granular fuel reactor (HTGR), the distribution is wide (close to Levy), and the $a$=2 model is important for calculations. In the case of an accidental core meltdown, the distribution has a power law, and the $a$=2 model is critical for safety. In routine engineering calculations of WWER reactors, power-law distributions are not used, since water “kills” all anomalies, converting them to standard diffusion. However, in more detailed studies (as in Ref. [15]) these models are used to understand the limits of robustness and subtle effects during run-time when statistics become 'jagged'.

In traditional kinetics (point kinetics equations), fluctuations are considered Gaussian. However, near criticality, the reactor behaves like a system undergoing a phase transition. In this case, neutron clusters develop as percolation paths, and the time to reach the emergency threshold L obeys this power law, not an exponential one. Fundamental scaling laws for the density of active sites and the survival probability $P(t)\sim t^{-\delta}$ are described in Ref. [12].

If the number of secondary neutrons or the mean free path has heavy tails, the traditional diffusion equation is replaced by the fractional diffusion equation Refs. [29, 30]. It has been shown that in highly inhomogeneous media or media with a fractal structure (e.g., porous absorbers), neutrons move not as particles in a gas, but as "Lévy jumpers."

## 4. Finite reactor volume

The finiteness of the reactor's size (L) imposes a truncation effect on Lévy flights. In the context of first-passage times (FPT), this truncation ensures that, over extended timescales, the distribution will asymptotically transition to an exponential form. Nonetheless, it is important to note that critical thresholds or hazardous conditions typically arise within shorter timescales, during which the impact of truncation remains negligible. Within this high-risk domain characterized by rapid accelerations, the reactor exhibits dynamics consistent with those of a "pure" directed percolation system, retaining all the associated anomalies inherent to such systems.

Let us write the formula for the truncated Levy distribution. Unlike the pure power law $P(k) \sim k^{-(1+a)}$ (9), $a$=1+α, which leads to divergence of moments (infinite variance), the truncated distribution introduces an exponential damping factor related to the reactor geometry (L) and the physics of capture:

$$P(k) \approx N \cdot k^{-(1+\alpha)} \cdot \exp(-\lambda k), \quad a=1+\alpha, \tag{10}$$

where α is the stability index (for the case $a$=2 we have α=1), λ~1/L is the truncation parameter determined by the characteristic size of the active zone or the path length to capture, N is the normalization coefficient.

The connection with the Levy-Khinchin formula here manifests itself in the modification of the logarithm of the characteristic function lnΦ(q). Instead of pure $|q|^{\alpha}$, which yields DP correlations, we obtain:

$$\ln \Phi(q) \sim const((\lambda^2 + q^2)^{\alpha/2} \cos(\alpha \arctan|q|\lambda) - \lambda^{\alpha}) . \tag{11}$$

Estimates show that the most dangerous situation for a WWER reactor is collective excitation within a single fuel assembly. Individual fluctuations in fuel elements are safe due to high λ. Global fluctuations of the entire reactor are suppressed by the Doppler effect and its enormous size. The "vulnerability window" is located precisely at the scale of a single fuel assembly (20-30 cm), where directed percolation is already effective and geometric truncation is still weak. The correlation length of noise is considered in Ref. [7]. Upon reaching the instability threshold, the correlation scale becomes comparable to the fuel assembly size (L=25 cm).

It can be concluded that the WWER geometric structure creates a hierarchy of stochastic risks. While percolation effects are suppressed at the microlevel (fuel rods) (λ≈1, λ≈1/L), at the mesolevel (fuel assembly) (λ≈0.04, L=25), the statistics of directed percolation with a power-law exponent $a$=2 can lead to abnormally fast local transient processes, requiring special attention when setting up core monitoring. A factor of L=25 maximizes the effective rate of stochastic runaway, reducing the time <T> to an accident by tens of percent.

The truncation parameter λ (or scale L) is most dangerous at intermediate intervals (the fuel assembly scale). At $a$=3, the system is local and geometry-insensitive. A neutron "doesn't see" the geometry of the entire assembly; it sees only its nearest neighbors. The system becomes local. At $a$=2, a resonance of stochastics and size occurs at the fuel assembly scale. However, at the scale of the entire core, global feedbacks suppress percolation effects, forcing the system back into the region of effective Gaussian distributions.

## 5. FPT functional, Lundberg equation and canonical form CF of stable distributions

Expressions for the moments of the FPT value are written in Ref. [9]. The relations for continuous and semi-continuous (top and bottom) processes differ in pre-exponential factors before the exponential; an expression for the MGF $T(s,x) = \boldsymbol{E}[e^{-s\tau^+(x)}, \tau^+(x) < \infty]$ of the form was obtained:

$$T(s,x) = \boldsymbol{E}[e^{-s\tau^+(x)}, \tau^+(x) < \infty] = e^{-\rho_+(s)x}, \; x > 0 , \tag{12}$$

$$\boldsymbol{E}[\tau^+(x), s] = -\frac{\partial \ln \boldsymbol{E}[e^{-s\tau^+(x)}, \, \tau^+(x) < \infty]}{\partial s} = \frac{\boldsymbol{E}[\tau^+(x) e^{-s\tau^+(x)}, \, \tau^+(x) < \infty]}{\boldsymbol{E}[e^{-s\tau^+(x)}, \, \tau^+(x) < \infty]} = x \frac{\partial \rho_+(s)}{\partial s} .$$

The calculations carried out in Ref. [9] show that there is practically no difference between expressions with and without pre-exponential factors, as in (12).

In Ref. [9], the expressions for the boundary functionals (12) include the roots of the Lundberg equation (8), (13). The Lundberg equation (k(r) from (8)) k(r)r=k=G(k)=s is the search for such a value $k(s)=\rho_+(s)$ at which the process generator (SCGF (7)-(8)) balances the Laplace transform parameter over time s. The Lundberg equation (13) translates events at the microlevel into seconds of time before the accident (macrolevel);

$$G(k)=s, \tag{13}$$

B (13) G(k) is the logarithm of the characteristic function (more precisely, the moment-generating function's (MGF) (4) $M(k)=E[e^{kX}]$, obtained from the characteristic function (6) by analytical continuation and replacement $iq \to k$), and s is the Laplace transform parameter for time. For the case $a$=2 (critical DP), the Levy-Khinchin integral is calculated analytically, leading to a logarithmic structure of the logarithm of the MGF $\Psi(q) = -\sigma|q|(1 + i\beta(2/\pi)\operatorname{sgn}(q)\ln|q|) + i\mu q$ Ref. [16]. This explicit form, corresponding to the critical $\alpha$=1, transforms the Lundberg equation (13) Ref. [9] into a logarithmically "clamped" asymptotic, explaining the anomalously high-power peaks. The Lundberg equation for the case of stable Levy distributions ($a\approx2$, $\alpha\approx1$) is transcendental. Its solution k(s) (notation $\rho_+(s)$ Ref. [9] in (12)) determines the behavior of the boundary functionals Ref. [9] (FPT, maximum of a random process, etc.). In the case of directed heavy-tailed percolation, we do not have a simple quadratic solution, as in ordinary diffusion ($Dk(s)^2=s \to k(s)=\sqrt{s/D}$).

In Ref. [16] it is noted that there are many different forms of recording the characteristic functions of stable laws, special cases of the canonical form (5). Five of them are given in Ref. [16]. In all five (as in other forms) there is a power dependence on the argument of the characteristic function, corresponding to the power dependence of the stable distribution function of the form (9).

For a stable law with index $\alpha\in(0.2]$, $a\in(1.3]$, the logarithm of the characteristic function in canonical form (A) [16] looks like this (14), (17):

$$\ln\Phi(q) = i\gamma_1 q - \sigma|q|^{\alpha-1}[|q| - \frac{iq}{|q|^{\alpha-1}}\omega(q,\alpha)],\ \ \omega(q,\alpha) = |q|^{\alpha-1}\beta tg(\frac{\pi\alpha}{2})\ \ (\alpha\neq1),\ \ \omega(q,\alpha) = -\beta\frac{2}{\pi}\log|q|\ \ (\alpha=1), \tag{14}$$

where $\gamma_1$ is the shift parameter (drift), $\sigma$ is the scale parameter, $\beta\in[-1,1]$ is the asymmetry parameter; for maximum asymmetry $\beta$=1, which corresponds only to positive power surges.

In problems on boundary functionals (FPT, maximum, etc.), which are considered in Ref. [9], we pass from characteristic functions (CF) (6) (Fourier transform with iq, CF (6)) to moment generating functions (4) (Laplace transform, parameter p or k). A change of variable is carried out. In the Lundberg equation, the argument k is often considered as a real parameter of the Laplace transform. This is equivalent to substituting $q \to -ik$ into the Fourier formula.

Asymmetry and positivity are important. We model the number of neutrons or the power ($\Phi\geq0$), the process is one-sided (strictly positive jumps). For such processes in Lévy's theory, the asymmetry parameter $\beta$=1 (maximum asymmetry). The imaginary parts in the expression from Ref. [16] are canceled or transferred to the real domain under certain transformations, describing a purely decaying or growing process. After the transition to the real form G(k) for strictly positive stable processes with index $\alpha<1$, the logarithm of the Laplace transform has a very simple form without imaginary units:

$$G(k)=\ln E[e^{kX}]=-\sigma k^{\alpha}. \tag{15}$$

This form is used in physics applications, as it describes the probability of chain survival. For the case $a$=2 ($\alpha\to1$), if the process is symmetric $G(k) \sim -|k|^{\alpha}$ (there is no imaginary part). If there is directionality (drift v), the term vk appears.

In the Ref. [9] (and in the theory of branching processes Ref. [26]), G(k) in (15) is understood to be a cumulant function: $G(k) = \ln \sum P(n) e^{-kn}$ . For power-law distributions, $P(n) \sim n^{-(1+\alpha)}$ (9) this integral (or sum) in the limit of small k gives precisely the power-law function $k^{\alpha}$.

In Ref. [16], the characteristic function (6) of the stable law $\Phi(t)=E[e^{itX}]$ has the form (форма (C) в Ref. [16]):

$$\ln \Phi(t) = -\sigma |-t|^{\alpha} \exp[-i \frac{\pi\alpha\theta}{2} sign(t)], \; 0 < \alpha \le 2, \; |\theta| \le \theta_{\alpha} = \min(1, \frac{2}{\alpha} - 1). \quad (16)$$

In the Lundberg equation in Ref. [9] (and in Ref. [26]), we work with the MGF $M(k)=E[e^{kX}]$ (4); at t=-ik:

$$\ln M(k) = -\sigma |-ik|^{\alpha} \exp[-i \frac{\pi\alpha\theta}{2} sign(-ik)] = -\sigma |k|^{\alpha} \exp[-\frac{\pi\alpha\theta}{2} sign(k)].$$

Since $|-ik| = k$ , and sign(-ik)=-i (in the complex plane), after transforming we obtain a real result for $G(k)=\ln M(k)$.

To ensure that the formula corresponds to the canonical form for one-way processes, the theory of stable laws uses the cosine relationship. For a strict transition (taking into account the normalization of σ in Ref. [16]): $G(k)=\ln M(k) \sim C(a) k^{\alpha}$. For k<0, for the growth of the system, this becomes $-\sigma_{eff} |k|^{\alpha}$ (Appendix A).

The imaginary unit in the Fourier transform describes the phase shift. By transitioning to the Laplace transform (t=-ik), we "rotate" the solution from the oscillation (wave) region to the exponential growth/decay region. Since our process (power) physically cannot be negative, all imaginary components must cancel out during this rotation. If the imaginary unit *i* remained, this would mean that the probability of reaching the level is a complex number, which is absurd.

As $\alpha \to 1$ (case $a$=2), the ratio $C(a) = |\Gamma(2-a)\cos((a-1)\pi/2)|$ obtained by integrating the power-law distribution of the process jumps turns into the uncertainty $\propto 0$. When expanded through the Γ-functions (as in the more general Levy-Khinchin form Ref. [16]), this coefficient yields π/2, and $G(k)=vk-\sigma(\pi/2)|k|^{\alpha}$.

The Lundberg equation is often written for negative values of the argument in the exponential (exp(-kX)), since this guarantees convergence for positive values of X. If we define $G(k)=\ln E[e^{kX}]$, then for a stable process with one-sided jumps at k<0 (exponential decay): G(k) behaves like $C(-k)^{\alpha}$. The sign in front of σ depends on whether we consider the "incoming" of neutrons (multiplication) or the "outgoing" (absorption/diffusion).

The generator G(k) in equation (13) is the logarithm of the MGF of form (4). For asymmetric processes (power growth), in the case of using form (A) (14) Ref. [16], it is written as:

$$G(k)=vk+\sigma C(a)|k|^{a-1}[1+\beta \cdot \mathrm{sign}(k) \tan(\pi(a-1)/2)], \; a \neq 2; \;\; G(k)=vk+\sigma C(a)|k|[k^{a-2}-\beta \cdot 2\log|k|/\pi], \; a=2, \quad (17)$$

$$\beta=0, \;\; G(k) = vk + \sigma C(a)|k|^{a-1}, \qquad C(a) = |\Gamma(2-a)\cos((a-1)\pi/2)|,$$

where $v$ is the deterministic drift (effective reactivity), the parameter $v$ (effective drift velocity) is the classical deterministic rate of power change. Physical meaning: it is determined by the excess reactivity $\rho$ and the average lifetime of a neutron generation $\Lambda$. Formula: $v \approx (\rho-\beta)\Phi_0/\Lambda$, where $\Phi_0$ is the initial flux; $G(k)=vk+V_{stoch}$. We take into account the Doppler effect by the effective contribution $v=v_{dr}-v_D$, $v_{dr} \approx (\rho-\beta)\Phi_0/\Lambda$, $v_D$ is the contribution of the Doppler effect directed against $v_{dr}$. The second term $V_{stoch}$ is the stochastic contribution $V_{stoch}=\sigma \cdot C(a) \cdot k^{a-1}$, β=0; in the general case, this is obtained in Section 7, when
$V_{stoch}(a, \beta, L)=\sigma \cdot \mathscr{R}(a, L) \cdot \Psi(a, \beta)$, $\mathscr{R}(a, L)=L^{\delta(a}$, $\Psi(a, \beta)=\Gamma(2-a)\,[|\cos(\pi(a-1)/2)|+\beta \sin(\pi(a-1)/2)]$.

DP theory and Lévy distributions are most relevant for HTGR, where fuel in the form of thousands of beads (TRISO) is loaded into the core. This is a classic problem of percolation in a random environment. If the beads are unevenly distributed, "flow paths" for neutrons may arise that are not accounted for by homogeneous codes. In MSR (Molten Salt Reactors), the fuel circulates. Directed percolation is complicated here by the fact that the "nodes" (fissile nuclei) themselves move through space. A relationship arises between flow turbulence and directed neutron percolation. Here, the Lévy-Khinchine formula describes the statistics of neutron transport in a turbulent flow.

What safety issues arise during "neutron bursts?" The main problem with systems with $a = 2$ is the unrepresentativeness of the average. This creates a control problem. In a conventional reactor, sensors at several points provide an adequate picture. In a system prone to DP with a power-law distribution, "local criticality" is possible. The power in one part of the core can increase 100-fold due to the random coincidence of "hubs," while sensors in another part of the core will not notice this due to the anisotropy of the process. The difference with the Gaussian description also lies in the accident statistics. In systems with a Lévy-Khinchin distribution (heavy tails), the probability of extreme events ("black swans") is orders of magnitude higher than predicted by a normal distribution. If the protection calculations are based on a Gaussian distribution, and the real environment, due to inhomogeneities, behaves like a DP system with $a$=2, then an "impossible" accident becomes statistically inevitable over a long period of time.

How it all works together. Geometry (DP) defines the framework along which neutrons can "run" (especially in inhomogeneous media). Statistics ($a$=2, Levy-Khinchin) predicts the possibility of giant jumps and localized bursts that defy average values. Feedbacks attempt to suppress these bursts (Doppler) or, conversely, inflate them (vapor coefficient).

A rigorous approach requires a transition from the diffusion approximation to fractional diffusion transport where the active zone is stochastic.

## 6. Neutron processes in the reactor and critical indices of the DP

The main conclusion can be formulated: although in normal modes the WWER is "protected" by water from fractal anomalies, understanding the directional percolation is crucial for startup modes (and some other situations noted above), where classical Gaussian statistics ceases to work. In Ref. [15] percolation relations of the form are used:

$$\xi \sim |r|^{-\nu} \quad \text{and} \quad \tau \sim \xi^{z} / D \sim |r|^{-z\nu} \tag{18}$$

for the correlation length $\xi$ and the diffusion time scale $\tau$. The values of the indices $\nu$=0.584 and z=1.901 calculated for the DP in Ref. [10] and used in Ref. [15] differ from the mean field values $\nu$=0.5 and z=2. This small difference in the index z has significant physical consequences. Thus, the diffusion equation for neutron transport is transformed into an equation with fractional derivatives, which leads to nonlocality on the reactor-wide scale. The dynamic critical index z ($z_{dyn}$) in (18) determines the rate of reaction moderation near the phase transition. It relates the relaxation time $\tau$ to the correlation length by $\tau \sim \xi^{z}$. The value of z depends on the universal class of the model (e.g., for the Ising model with conservative dynamics $z \approx 2+\alpha/\nu$ or about 2). Most often, $z \approx 2$ (for diffusion processes) or $z \approx 2-\eta \approx 2$ (for non-conservative dynamics in some models). The index z indicates how much the relaxation time increases as the transition point is approached ($\xi \to \infty$). It is calculated using the renormalization group or Monte Carlo methods. A precise value requires a model specification. The index $z$ determines the coupling between space and time ($\xi_{\parallel} \sim \xi^{z}_{\perp}$), as well as the propagation velocity of disturbances in the core. Even if $z$ varies, the topological distance (chemical index) between generations remains equal to 1 by definition. However, the average path in such graphs (the small-world effect) grows as $l \sim \ln(N)$ for finite-variance distributions, $l \sim \ln(\ln(N))$ for scale-free trees ("ultra-small world"). In DP, two different correlation radii appear:

$\xi_{\perp} \sim |p - p_c|^{-\nu_{\perp}}$, (across) and $\xi_{II} \sim |p - p_c|^{-\nu_{II}}$ (along the direction of neutron motion),

where p is the probability of nuclear fission by a neutron. Effective neutron multiplication factor $k_{ef} = p\bar{\nu}$, where $\bar{\nu}$ is the mathematical expectation of the number of secondary neutrons in one fission event; to the case $k_{ef}=1$, $p_c=1/\bar{\nu}$ .

Fractal theory Refs. [15, 18-21] is most important for compact thermal systems (research reactors, small modular reactors). In large-scale fast reactors, classical physics works more reliably, since the "size of a neutron" (its range) is too large to become entangled in fractal labyrinths.

The application of DP to neutron propagation in a reactor at $D$=2.4 ($D$ is the fractal dimension) is entirely valid. This value practically coincides with the theoretical fractal dimension of the active DP cluster in three-dimensional space ($d_f$≈2.46). This description allows us to consider the critical state of the reactor as a self-organizing criticality, where the flow geometry (fractal) and the reaction time (index $z$) are rigidly related by universal scaling laws. Using this pair of values ($D$=2.46 and $z$=1.901) is a physically and mathematically consistent solution, as shown in Refs. [15, 24, 25]. This allows us to construct a rigorous model of neutron fields within the framework of the DP theory for three-dimensional space ($d$=3). In this case, the consistency condition for the dimension $D$=2.46 is satisfied. In the directed percolation theory for the (3+1)-dimensional case (3 spatial dimensions + 1 time), the fractal dimension of the critical cluster $df$ is calculated using the standard critical indices: $D = d - \beta\nu_{\perp}$. For 3D space, numerical estimates give $D$≈2.46...2.48. This value is ideally suited to justify the "branching" of neutron chains mentioned by Mandelbrot Ref. [20] (that for trees with infinitely thin trunks, the fractal dimension D serves as a measure of branching) (the original value $D$=2.4 was a rounding, and 2.46 was a refinement for the pure DP model). The index $z$ (characterizing scale anisotropy) relates spatial and temporal correlations: $\xi_{\parallel} \sim \xi^{z}_{\perp}$. In the context of a nuclear reactor, $\xi_{\perp}$ is the characteristic size of the neutron flux "spot" in the core (spatial coherence), $\xi_{\parallel}$ is the lifetime of a neutron colony or the system response delay. The value $z$≈1.901 indicates that the relaxation time of disturbances in the reactor grows almost as the square of its geometric size ($t \sim L^{1.9}$), which is close to classical diffusion ($z$=2), but takes into account fractal "delays" and branching effects. We will provide a physical interpretation for a reactor. Using these constants, we can state the following: 1. Criticality is a phase transition: the reactor is exactly at the point where neutron cascades become self-similar fractals. 2. Fission topology: the number of secondary neutrons $\bar{\nu}$ ≈2.4 determines the space filling density of the fission cascade. 3. Anomalous kinetics: since $z$<2, the transfer of information (or power disturbances) in the reactor occurs somewhat faster than predicted by classical diffusion theory for a homogeneous medium, due to the "long tails" of the range distribution in a highly inhomogeneous medium. The value $D$=2.46 characterizes the global topology of the neutron tree, but the local fissions density exhibits multifractal properties Ref. [21]. This is due to the spatial segregation of moderation and breeding processes in the inhomogeneous WWER lattice. Negative feedback loops (the Doppler effect) act as a nonlinear filtering mechanism, limiting the multifractal dimension of the most powerful fluctuations, thereby ensuring the nuclear safety of the system. The dimension $D$=2.4 serves as a scaling exponent relating the number of particles (neutrons) to the spatial extent of the reaction zone, differing from the Euclidean dimension $d$=3 due to the complex, non-compact nature of diffusion paths. The DP describes the mechanism of instantaneous propagation (kinetics). Feedback loops limit the growth of the fractal, preventing it from turning into an infinite cluster (explosion). As a result, feedback loops "cut off" the multifractal tails of the distribution. This justifies the use of the limited percolation model, where $D$=2.4 is the starting point, which is in fact stabilized by thermodynamics. Thus, the set of critical indices proposed in Ref. [15] ($D$=2.46, $z$=1.901) allows us to move from empirical branching estimates to a rigorous statistical description of the spatio-temporal kinetics of a reactor as a system in a state of self-organized criticality (SOC). The Doppler effect "cuts off" fluctuations. In multifractal terminology, this means that feedback loops narrow the multifractal spectrum [21]. In a "pure" DP process, the spectrum is broad (extremely strong localized peaks are possible). In an operating reactor, feedback loops suppress the sharpest peaks (singularities), making the system more homogeneous (monofractal). Using the value $D$=2.4 is an excellent approximation that takes into account the "looseness" of the neutron tree in a real heterogeneous environment (e.g., in a WWER), where part of the space is occupied by structural materials. Following Mandelbrot Ref. [20], the fractal dimension is related to the number of secondary neutrons ($\bar{\nu}$ ≈2.4 for $^{235}$U) in the Euclidean space $R^3$. This is the dimension of a real trajectory in physical space, where scattering angles and free paths are taken into account. The difference in dimensions is due to the transition from a topological description of the multiplicative process (Cayley tree, D=4), which takes into account only branching logic, to a geometric description in the reactor's phase space (D=2.4), where branching is limited by the physical volume and transport conditions in the water-water medium. The DP model is adequate for describing a real reactor as a physical object in space. Why is DP better suited for a reactor? Spatial availability: in a reactor, neutrons don't simply "split," they diffuse in three-dimensional space. DP takes into account that "descendants" can end up in the same region of space as their "ancestors." Excluded volume (or saturation) effect: DP takes into account that a node cannot "split" indefinitely (cores burn out, or the flux becomes limited). Anisotropy ($\nu_\perp$ vs $\nu_{\text{II}}$ ): $\nu_{\text{II}}$ (a long time) describes how long a deviation from criticality lasts, $\nu_\perp$ (in space) describes the critical radius (core size). In a reactor, this is critical: if the core size is smaller $\xi_\perp$, percolation (chain reaction) will not become self-sustaining, even if $p>p_c$.

Feedback and Index Changes. In a real reactor, negative feedbacks exist (e.g., Doppler broadening of resonances during heating or an iodine well). From the perspective of phase transition theory, this transforms the system from "ordinary" directed percolation to a system with self-organized criticality (SOC) or to the class of DPs with absorbing states. Index Changes: If the feedback is instantaneous and strong, its "locks" the system near the threshold. This can change the effective index (the power of an infinite cluster). Instead of a smooth increase, we can observe jumpy or oscillatory regimes. Diffusion Scale: In DPs, the diffusion time is related to space through the dynamic index $z_{dyn}=\nu_{\text{II}}/\nu_\perp$. For the mean field, z=2. If anomalous transport occurs in the reactor (e.g., via fast, long-range neutrons), this dynamic index changes. DPs and Feedback Loops. For physicists, DPs are a powerful and important tool because they take into account the spatial correlation length. It is through this correlation that the relationship between the geometric dimensions of the reactor and its operating time (the operating period of the reactor) is expressed.

## 7. FPT calculation

The numerical order of magnitude must correspond to the actual physics of noise in the active zone (frequency range 0.1-10 Hz). Since we are working with relative power (Φ is dimensionless), the dimension of σ must be [$cm^{-1}$ $s^{-1}$].

Here is an example of physically substantiated values for fuel assemblies (WWER or BWR types). Input parameters (Case Study): fuel assembly scale (L) 25 cm (characteristic size of the vapor correlation region); emergency stop setpoint ($\Phi=\Delta P/P_0$) 0.1 (10% power excess); deterministic drift (v) 0.05 $s^{-1}$ (slow reactivity growth); Doppler effect ($v_D$) -0.02 $s^{-1}$ (compensation); effective drift ($v_{eff}=v-v_D$) 0.03 $s^{-1}$; numerical value of σ: for reactor noise in the developed boiling mode, the fluctuation intensity is usually a small fraction of the nominal value per unit length: σ=0.002 ($cm^{-1}$ $s^{-1}$).

Calculations were performed in accordance with five different forms of the logarithm of the characteristic function given in Ref. [16]. In some cases, the average time to reach the level becomes negative. Mathematically, this means that the "anomalous jumps" against the current (due to the structure of the form of the logarithm of the characteristic function and the sign) have become stronger than the directional drift, and the cluster, on average, "moves away" from the barrier rather than approaches it. For example, this was obtained for a change in the parameter a in the range of 1.25-1.55 of the form (M) of the form (19) from Ref. [16]:

$$G(k)=\lambda[k\gamma+|k|^{\alpha}+k\omega_M(k,\alpha,\beta)],$$

$$\omega_M(k,\alpha,\beta)=(|k|^{\alpha-1}-1)\beta tg(\alpha\pi/2),\ \alpha\neq 1;\ \ \omega_M(k,\alpha,\beta)=-\beta(2/\pi)\log|k|,\ \alpha=1. \quad (19)$$

As obtained in expression (12), the average time <T> is related to the derivative of the generator G`(k); differentiating both parts of equation (13) with respect to s, we find: $G'(k)$ dk(s)/ds=1, dk(s)/ds=1/$G'(k)$,

$$\langle T\rangle=\Phi_{crit}\,\mathrm{dk(s)/ds}=\Phi_{crit}/G'(k). \quad (20)$$

If at β=0, $G(k)=vk+\sigma C(a)|k|^{a-1}$ (17), then its derivative:

$$G'(k)=v+\sigma C(a)(a-1)|k|^{a-2}. \quad (21)$$

In a more general case, we write the expression with $\Psi(a,\beta)$ and $L^{\delta(a)}$:

$$G(k)=vk+V_{st},\quad \Psi(a,\beta)=[C(a)+\beta\Gamma(2\text{-}a)\cdot\sin(\pi(a-1)/2)],$$

$$\langle T\rangle(a)=\Phi_{crit}/(v+V'_{st}),\quad V'_{st}=\sigma\cdot(a\text{-}1)L^{\delta(a)}\,\Psi(a,\beta).$$

Let us clarify this notation. In (21), the substitution k=1/L is made, the exponent -($a$-2) is denoted as δ($a$). Singularities at $a$=2 and $a$=1 arise when s=0 and k(s=0)=0. But in Refs. [22, 23] (as well as in the Introduction), it is noted that the “field” s, conjugate to the FPT (in Refs. [22, 23], this “field” is denoted by $\gamma$), is equal to zero only in equilibrium, while we are considering a more general nonequilibrium situation. The relation $s\sim 1/t$ is valid, and for $s\to 0$, $t\to\infty$. However, the time moment $t\sim 1/s$ must fall into the finite interval of the FPT that we are considering. Therefore, the values of s must also be greater than zero. Determining such a value of s is a complex problem. We will assume that temporal quantities are related to spatial quantities through some velocity and consider the spatial problem of finite reactor dimensions.

In addition to the physical limitation of the parameter $s\sim 1/t$, there is a physical limitation of the parameter k. In an infinite mathematical environment, k can tend to zero ($s\to 0$), which for $a$=2 yields an infinite "speed" of fluctuations. But a reactor is a finite system of size L. In the theory of waves and stochastic processes, there is a fundamental relationship: the minimum wave number k is inversely proportional to the maximum size of the system L: $k_{min}\approx 1/L$. This is a truncation; the system "does not feel" correlations longer than its own size. Now let us substitute $k\sim 1/L$ into the stochastic term $\sigma C(a)(a-1)(1/L)^{a-2}=\sigma C(a)(a-1)L^{2-a}$. The factor $L^{\delta(a)}$ was hidden in the constant "scale", but it is precisely this factor that is the "engine" of the anomalous behavior of the system during the transition to directed percolation ($a\to 2$).

In directed percolation (DP) and truncated Lévy flights theory, this factor describes finite-size scaling. It shows how the "weight" of a stochastic cluster increases with the size of the region it can "shoot through".

Here's how to mathematically correctly account for $L^{\delta(a)}$ and its properties. The exponent $\delta(a)$ is not a constant, but a function related to the fractal dimension of the neutron cluster. Based on the theory of anomalous diffusion and scaling in DP, it can be represented as follows: $\delta(a)=\max(0, 3-a)$. (Here we use the relationship with the stability index $\alpha=a-1$: in the Lévy regime, the effective cluster power grows as $L^{2-\alpha}\to L^{3-a}$).

Why do we use $\delta(a)=3-a$ in our calculations? This is where the spatial dimensionality comes into play. In a 3D reactor (WWER), a neutron cluster propagates not along a line, but through a volume. According to the theory of directed percolation in 3D, the effective contribution of fluctuations to the process "velocity" is scaled by an additional factor that accounts for volumetric connectivity.

In 1D (chain), $\delta=2-a$. In 3D (core), $\delta=(2-a)+(d-1)$, where d=2 (the effective dimension for DP). For directed percolation on the fuel assembly, this yields a resulting exponent of $\delta(a)=3-a$.

Thus, the formal expression is transformed into a physical one, and the expression $C(a)(a-1)k^{a-2}$ after replacing $k\sim 1/L$ and taking into account the 3D volume effect takes the form $\sigma C(a)(a-1)L^{3-a}$.

Let's analyze this situation using the example of $a$=2. Mathematically: for $\delta=(2-a)$, $(a-1)k^{a-2}=(2-1)k^{2-2}=1\ k^{0}=1$ (it would seem that the size of L is unimportant). However, physically, DP must be taken into account. For $a$=2 (the critical point), the cluster becomes fractal. Its "strength" grows linearly with the size of the region it captures. Therefore, $\delta(a)=3-a$ and the factor $L^{3-2=1}$=25 appears. This is what turns the "imperceptible" contribution of $\sigma$=0.01 into a noticeable jump 0.01 1.57 25 ≈ 0.39 (plus about 40% to the speed).

The factor $L^{\delta(a)}$ introduces geometric resonance into the Lundberg equation. As long as $a$>3, the reactor is "unaware" of its size (the processes are local). As soon as a drops to 2.0, the factor L transforms a small stochastic fluctuation $\sigma$ into a macroscopic power surge. A special feature of $a$=2 is that this is the point where the cluster's fractal dimension allows it to "sense" the fuel assembly boundaries. This is why the time curve <T> on the graph has a kink at this point.

The parameter $\lambda$ from expressions (10), (11) is the mathematical expression for the finite size of L. In the theory of truncated Lévy Flights, the parameter $\lambda$ is introduced into the logarithm of the characteristic function precisely to limit infinite jumps. Physically, this limit is set by the size of the fuel assembly. Thus, substituting $k\sim 1/L$ into the derivative G′(k) is the same as calculating G′(k) for a function that already contains $\lambda$, as in (10), (11). How $\lambda$ "generates" the factor $L^{\delta(a)}$. Let's look at the derivative of the generator with truncation: $G(k)=vk-\sigma[(\lambda-k)^{\alpha}-\lambda^{\alpha}]$. Let's take the derivative of $G'(k)$ at $k$=0 (which corresponds to finding the average time): $G'(0)=v+\sigma\alpha\lambda^{\alpha-1}$. Now let's substitute the physical meaning of $\lambda$=1/L and $\alpha=a-1$: $\sigma(a-1)(1/L)^{a-2}=\sigma(a-1)L^{2-a}$, we obtain expression (21).

Thus, $\lambda$ is the truncation mechanism (why the jumps are finite), and $L^{\delta(a)}$ is the magnitude of this truncation's contribution to the 3D system. There is one physical limitation (the size of the fuel assembly), which manifests itself in the Lundberg equation as the parameter λ, and in the final formula for time as a factor L raised to a power dependent on *a* and the spatial dimension.

When moving to the fuel assembly scale (L), the stochastic term is transformed to account for the system's dimensionality. The factor (*a*-1), which is part of the generator's derivative, approaches unity as $a \to 2$ and is incorporated into the effective intensity parameter σ, yielding the final form $\sigma C(a) L^{3-a}$.

In statistical mechanics, the Laplace transform (the generator G(k)) is closely related to the generating function of cumulants: $\psi(k)=\ln\langle e^{kX}\rangle$.

In thermodynamics of trajectories, Refs. [31-32] Ψ(k)=G(k) plays the role of dynamic free energy, which corresponds to the topological pressure of dynamic system theory. In equilibrium thermodynamics, k is the conjugate field (just as the inverse temperature β is conjugate to the energy E). In our problem, k is the parameter of the control field, which "selects" from all possible reactor trajectories those that lead to an emergency release. We can find critical values of the "field" k and the index *a*, at which the system undergoes a dynamic phase transition from a stable state to the state of a "ruined" (emergency) system.

We consider the logarithm of MGF as the potential of 'dynamical free energy' in the s-ensemble of trajectories Ref. [31–32]. The singularity of this potential at *a*=2 is the mathematical reason why the 30% increase in risk (from (17)) becomes inevitable.

At the point *a*=1, a singularity of the potential $\Psi(k) \to \propto$ itself occurs (structural catastrophe), the potential ceases to be smooth at k=0. In the theory of stable distributions Ref. [16], the condition *a*>1 (α>0) is valid. The neutron population can violate this condition. However, in finite systems, the equality k=0 is not satisfied. For *a*>1 (short-range regime), the system behaves as in ordinary directed percolation (DP). The long-range action decays quickly enough not to affect the critical exponents. For a~1 (long-range regime), "Lévy flights" become dominant. For *a*=2, the potential remains finite, but its first derivative becomes singular, which marks a change in the regime from short-range to a regime with dominant long-range action ("Lévy flights"), changing the universality class of the system. The system passes to a new regime, where the critical exponents begin to depend on the parameter *a*. A first-derivative singularity in this context denotes a jump or break in thermodynamic quantities or transport coefficients. This is a classic sign of a phase transition (often second-order or higher, according to Ehrenfest's classification), where the system changes its global response to small perturbations. This is a cross-phase transition in directed percolation models.

At the point of directed percolation (*a*=2) the following occurs: the curvature changes sign: up to *a*=2 the free energy Ψ(k) is convex, which corresponds to the central limit theorem (stability). At the point *a*=2 the curvature (analogous to specific heat) becomes infinite. Condensation of trajectories: neutron trajectories begin to "condense" into narrow, long clusters. This is analogous to the "gas-liquid" phase transition, where the "liquid" is the continuous chain of fission events throughout the entire fuel assembly. Logarithmic contribution: it is at this point that the correlation length ξ becomes equal to L due to critical fluctuations. The contribution to the denominator becomes σlnL. Ref. [15] proved that the critical point of a nuclear reactor corresponds to the phase transition of directed percolation.

The free energy of trajectories Ψ(k)=G(k) exhibits different types of singularities: at *a*=3, a singularity of the second derivative $\Psi''(k) \to \propto$ (a second-order phase transition) occurs. At *a*=2, a singularity of the first derivative of the potential $\Psi'(k) \to \propto$ occurs. A singularity of the first derivative is always a sharp release of energy or a sharp change in density. The finiteness of the system smooths out these effects.

The risks associated with underestimating local criticality are explored in Refs. [15, 24, 25] on "neutron clustering" (particle clustering). Studies of stochastic environments discuss the effect of unexpected criticality during density fluctuations, whereby a difference of 1–2% on average can translate into significant deviations in local fluctuations.

There are two possible ways to write the G(k) function. The difference in signs (plus or minus) and absolute values depends on the analytical continuation to the complex plane. The minus sign at β=0 ( $vk - \sigma C(a)|k|^{a-1}$ ) is typically used in diffusion problems (the Fokker-Planck fractal equation), where stochastics "blur" the concentration (reduce it at the center). The plus sign ( $vk + \sigma C(a)|k|^{a-1}$ ) is used in risk theory problems Ref. [9]. Here, stochastic jumps (Lévy flights) "help" the system reach the threshold faster. Since we are looking for the root of the G(k)=s equation, where s is the absorption rate, the plus sign physically means that random power surges shorten the time <T>. We are considering the most dangerous case.

We use a form that emphasizes the stability index: $G(k) = vk + \sigma C(a)|k|^{a-1}$. Here, k is taken positive, since we are seeking the root in the right half-plane of the Laplace transform. Stochastic "flights" in our model operate in the same direction as positive reactivity—they accelerate the attainment of the critical threshold. The physical meaning of the choice of the "plus" sign is simple: v is the average velocity of the upward power "drift" (the deterministic part). The second term is the contribution of stochastic "Lévy flights." Since the "flights" (power surges) are directed in the same direction as the drift, they increase the overall velocity approaching the threshold. The higher the velocity in the denominator (G`(k)), the shorter the time <T>. Stochastics "speed up" the process, reducing the safety time. Therefore, the denominator contains the sum.

The diversity of different forms of representation of stable distributions was noted above [16]. Accordingly, there are a significant number of expressions for the average time to reach some dangerous limit. We present some of them together with graphs constructed from these expressions.

In [16], the form (C) (16) was obtained, which is valid for strict stability, in the terminology of Feller [17]. The expression for the derivative of the cumulant generating function after using the exponent $\delta(a)=3-a$ has the form:

$$\langle T(a)\rangle = \Phi_{crit} / G`(k), \qquad G`(a) = v_{eff} + \sigma(a-1)L^{3-a}\exp[-(3-a)\pi/2],\ 1<a\leq 2;$$

$$G`(a) = v_{eff} + \sigma(a-1)L^{3-a}\exp[-(a-1)\pi/2],\ 2\leq a\leq 3. \qquad (22)$$

Figure 1, calculated by expression (22), with $v_{eff}$=0.03, σ=0.002, β=1, shows the dependence of the average time to reach <T>(*a*)=T(*a*) level $\Phi=\Delta P/P_0=0.1$ on the parameter *a*, for form (C) [16].

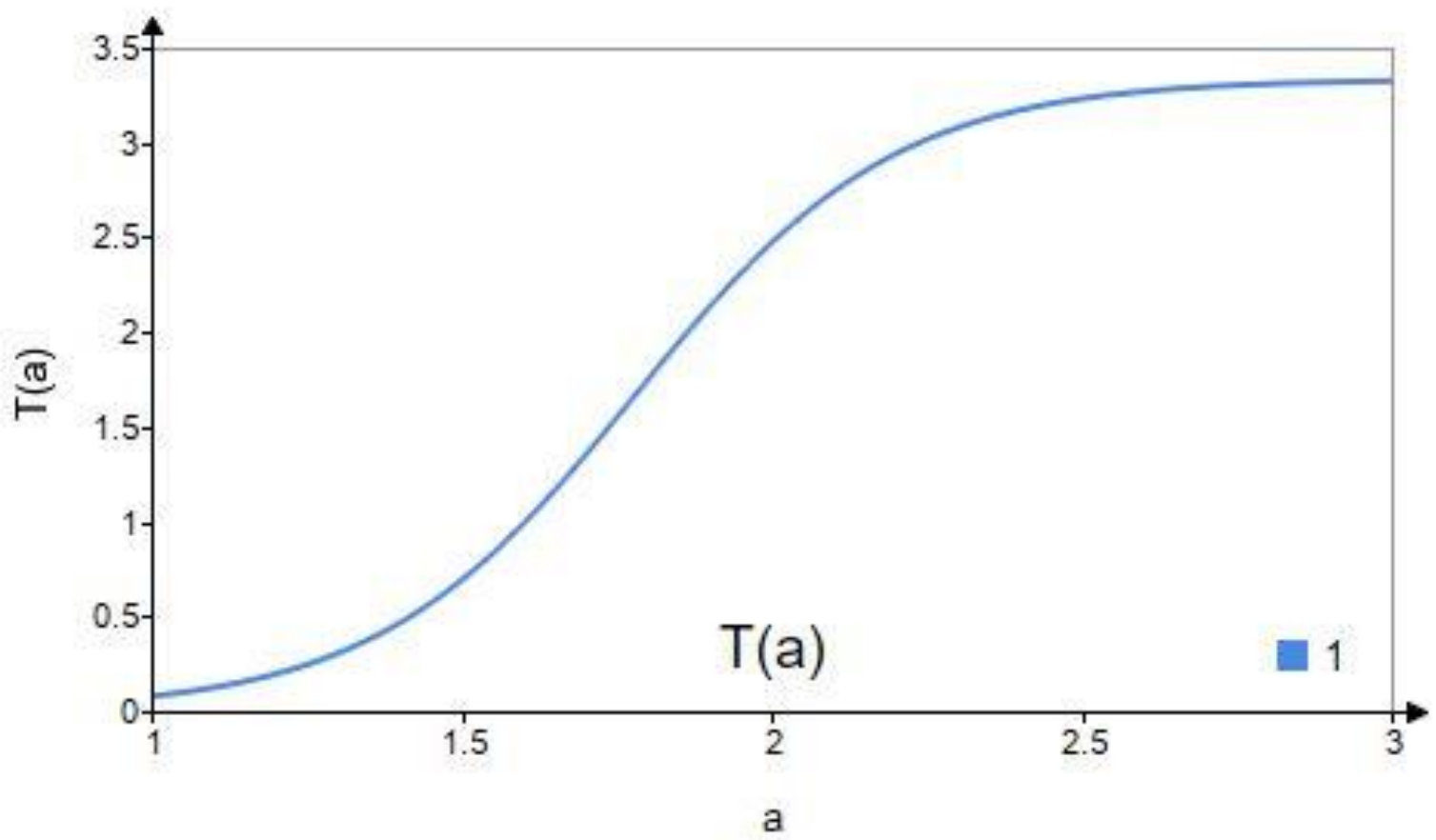

Fig. 1. Dependence of the average time to reach <T>(*a*)=T(*a*) the level $\Phi=\Delta P/P_0=0.1$ on the parameter *a*, for form (C) [16], calculated using formula (22), with $v_{eff}$=0.03, σ=0.002, β=1.

The dependence of the average time to reach <T(*a*)>=T(*a*) the level $\Phi_{crit}$ on the parameter *a*, calculated for form (A) [16] using expression (17) taking into account the Doppler effect, as noted after (17), with β=1, δ(*a*)=3-*a*, L=25, v= $v_{dr}$-$v_D$ =1-0.2=0.8 ($v_D$ =0/2) has the form:

$$\langle T(a)\rangle = \Phi_{crit} / G`(k),\ G`(k) = v + \sigma(a-1)L^{3-a}\Psi,\ \Psi = C(a) + \beta\Gamma(2-a)\sin(\pi(a-1/2)), \qquad (23)$$

Calculation of the dependence of the average time to reach <T>(*a*)=T(*a*) the level Φ on the parameter *a*, calculated using formula (23) for β=1, requires the elimination of features for *a*=2 and *a*=1.

From form (A) [16], after "smoothing" the features, the ratio T(*a*) is written as the dependence of the average time to reach the level on the parameter *a* in the form:

$$T(a) = \Phi / G`(a),\quad G`(a) = dG(k)/dk,\quad k = 1/L, \qquad (24)$$

$$G`(a) = v_{eff} + \sigma(a-1)L^{3-a}\Gamma(2-a)\cos((a-1)\pi/2)\sin(\beta(a-1)\pi/2)/\beta(a-1)\pi/2,$$

The behavior of the dependence of the form (24), where σ=0.002, $v_{eff}=v_{dr}-v_D=1-0.2=0.8$, L=25, is shown in Fig. 2. As the parameters change, the shape of the curve changes. The derivation of expression (24) is described in Appendix A.

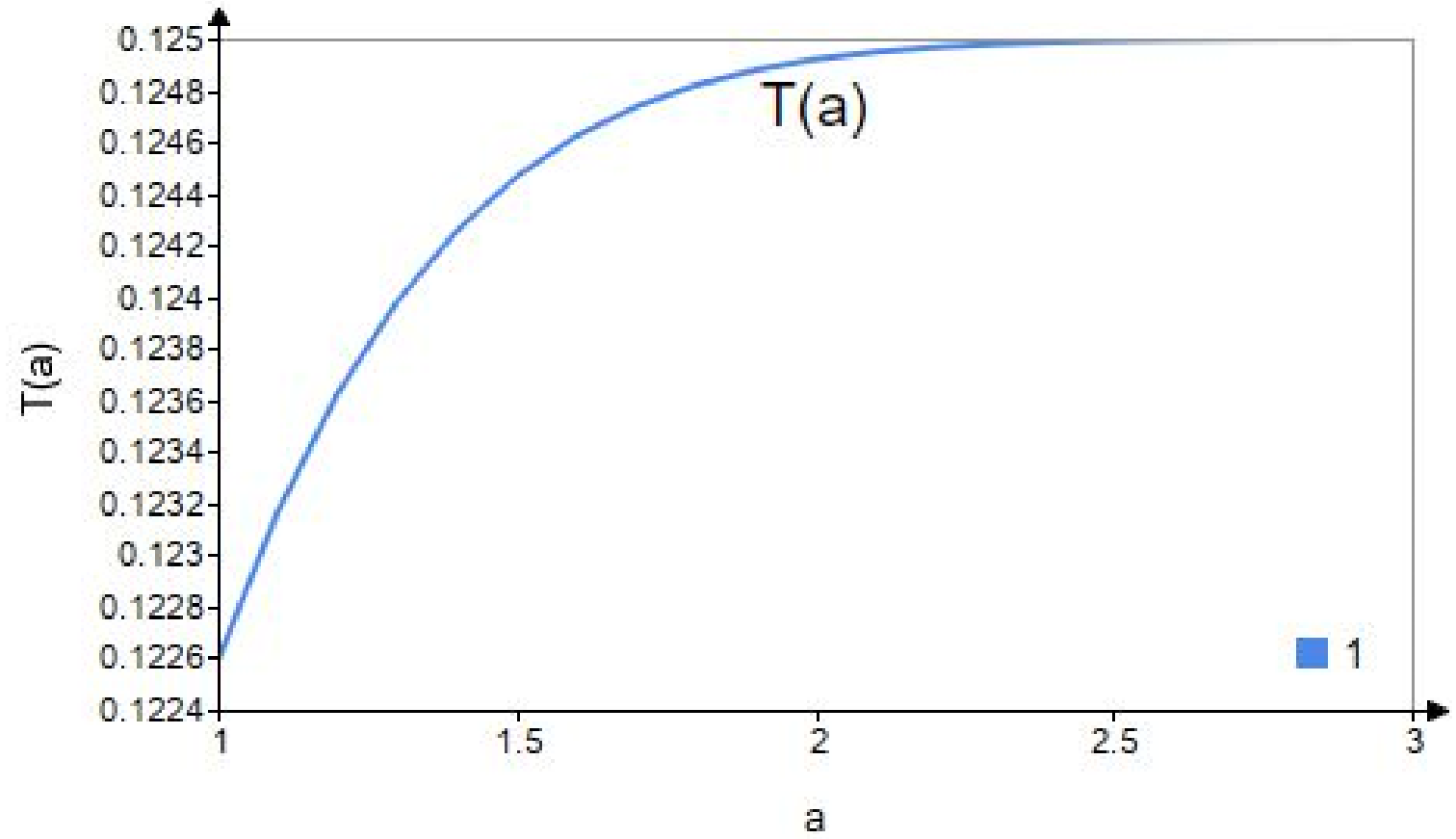

Fig. 2. Dependence of the average time to reach <T>(*a*)=T(*a*) the level Φ=Δ P/$P_0$=0.1 on the parameter *a*, for form (A) [16], calculated using formula (24), with $v_{eff}$=0.8, σ=0.002, δ(*a*)=3-*a*, β=1.

When changing the parameter σ from σ=0.002 to σ=$10^{-2}$, the appearance of the curve changes, Fig. 3.

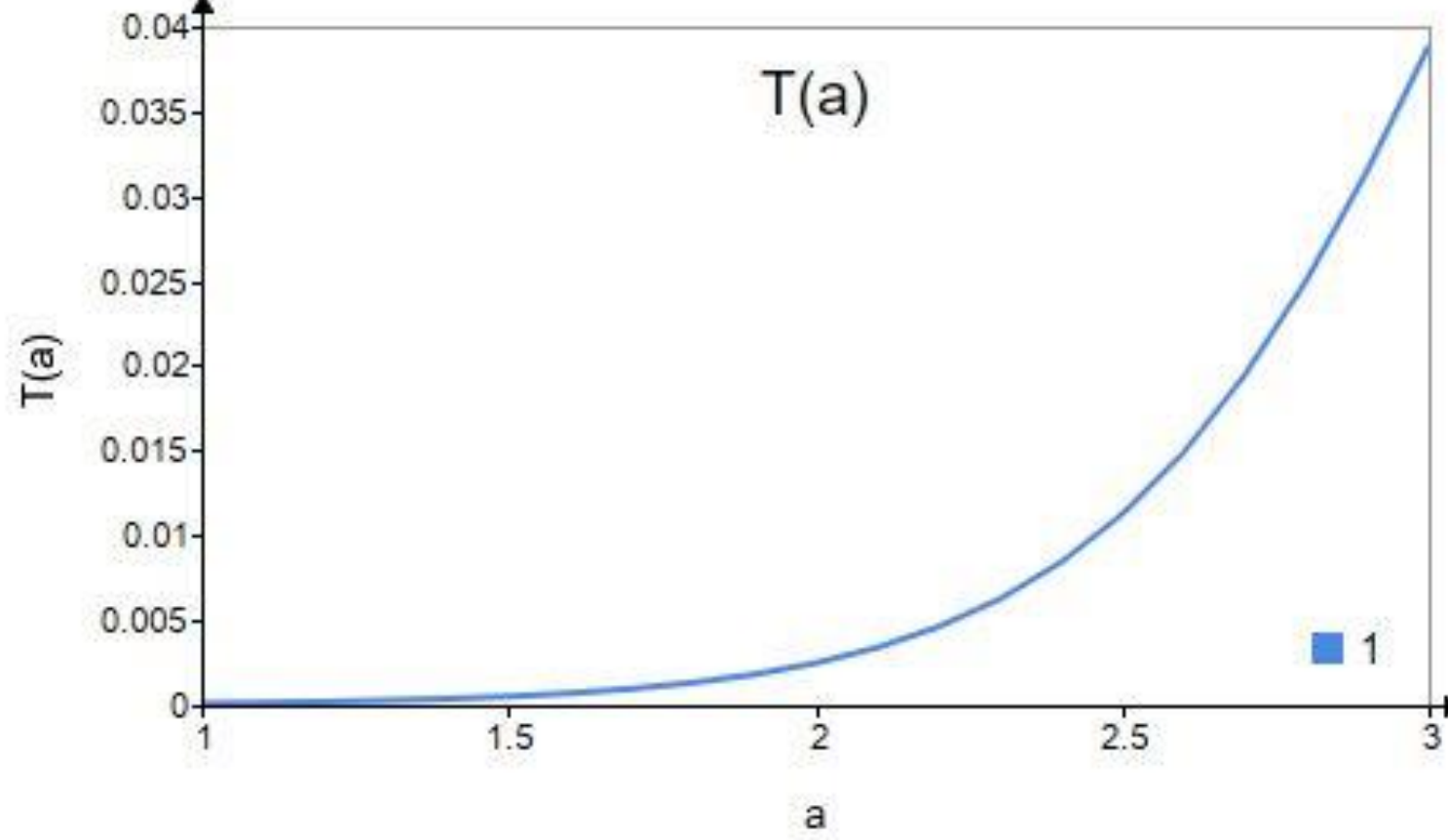

Fig. 3. Dependence of the average time to reach <T>(*a*)=T(*a*) the level Φ=Δ P/$P_0$=0.1 on the parameter *a*, for form (A) [16], calculated using formula (24), with $v_{eff}$=0.8, σ=$10^{-2}$, δ(*a*)=3-*a*, β=1, L=25.

With another approach to "smoothing" the features of form (A), an expression of the form is written:

$$G`(a) = v_{eff} - \sigma(3-a)L^{2-a}\cos(\Phi(a))\Gamma(3.047-a)/(2.047-a), \qquad (25)$$

$$\Phi(a) = \beta(3-a)(\pi/2)(1+2\ln(k)/\pi),$$

where σ=0.0932, $v_{eff}$=0.025, the singularity at *a*=2 in Fig. 3 is very approximately eliminated by replacing $\Gamma(2-a) \approx \Gamma(3.047-a)/(2.047-a)$. Fig. 4 shows the calculation using expression (25). For β=1, *a*=2, the uncertainty is revealed: the gamma function $\Gamma(2-a)_{a\to 2} \approx 1/(2-a)$, $\cos(\pi/2) \to 0$, $\cos(\pi(a-1)/2)\,\Gamma(2-a)_{a\to 2} \to \pi/2$.

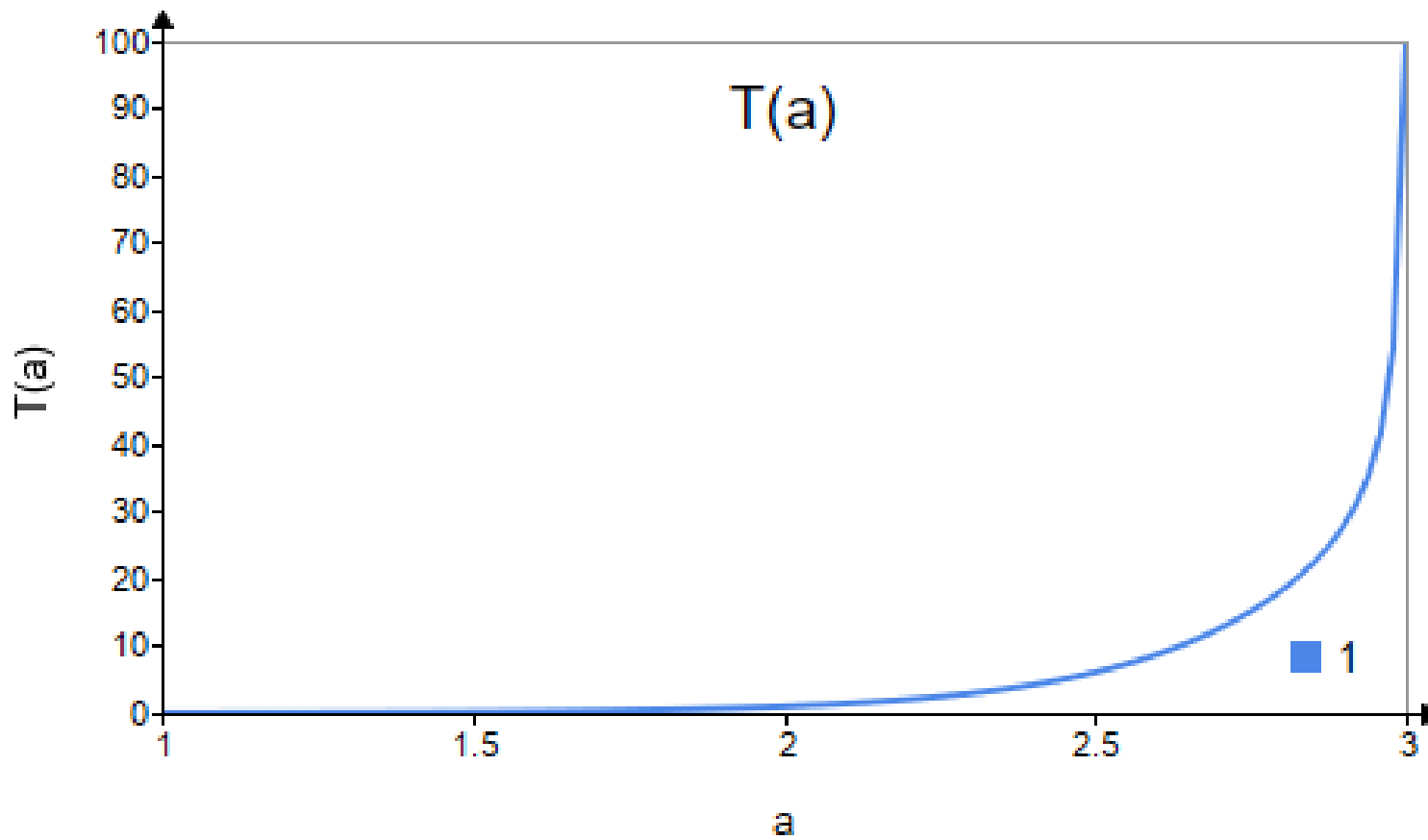

Fig. 4. Dependence of the average time to reach <T>($a$)=T($a$) the level Φ= Δ P/$P_0$=0.1 on the parameter $a$, for form (A) [16], calculated using formula (25), with $v_{eff}$=0.025, σ=0.0932, β=1, L=25.

Changing the parameters in (25) changes the curve's course. Thus, it "rises" at σ=0.0632, $v_{eff}$=0.025; changing the parameter σ changes the curve's appearance, Fig. 5.

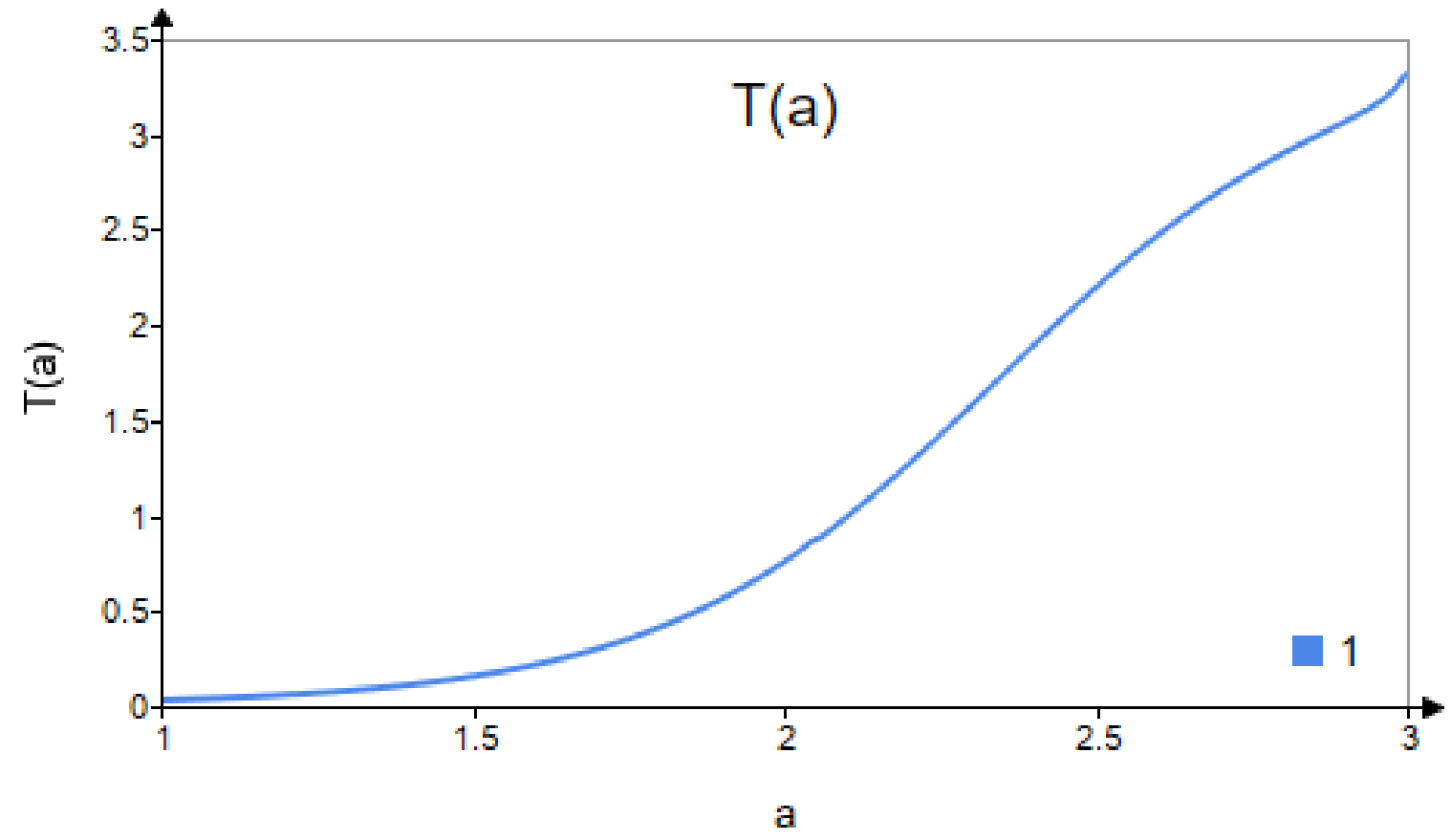

Fig. 5. Dependence of the average time to reach <T>($a$)=T($a$) the level Φ= Δ P/$P_0$=0.1 on the parameter $a$, for form (A) [16], calculated using formula (25), with $v_{eff}$=0.025, σ=0.0632, β=1, L=25.

**8. Conclusion**

The proposed approach to the theory of nuclear reactors has many aspects and provides many opportunities for a more detailed description.

The way neutrons move and spread in a nuclear reactor, along with how they form complex patterns, shows the complicated processes happening during nuclear fission and neutron movement. Using certain rules from percolation theory that apply to special types of lattices, we can understand how important factors, like the multiplication factor, change over time. These estimates show that getting exactly one multiplication factor can only happen if there are an infinite number of neutron generations, which means an infinite amount of time and an infinitely large reactor system. But for real-world conditions, we can figure out how long it takes to reach very small changes of multiplication factor.

These findings are important for reactor safety, especially in situations with fewer neutrons, like when a reactor is starting up or in critical assemblies. They also help in better understanding how reactors behave during sudden changes, transient processes in the reactor.

This paper analyzes the applicability of DP theory and fractional calculus to neutron field dynamics in nuclear reactors. Based on the material analyzed, the following conclusions can be drawn.

It has been established that the classical diffusion model is a special case of more general stochastic dynamics. At the critical point ($k_{eff} = 1$), the branching process of neutron fission naturally transitions to the universality class of directed percolation. When $a \le 2$, the standard Laplace operator in the transport equation is correctly replaced by a fractional operator generated by the canonical Lévy-Khinchine form.

Under certain conditions, a “patchy” effect (Patchy clusters) appears. Recent studies Refs. [15, 24, 25] and experiments on critical assemblies confirm that near the percolation threshold, the neutron field loses spatial homogeneity, forming fractal clusters. This phenomenon is a direct consequence of directed percolation in spacetime, where the system's “memory” is maintained by delayed neutrons.

In WWER nominal modes, the high density of the moderator (water) effectively suppresses the “heavy tails” of the distributions, limiting the discrepancy between the DP model and classical diffusion to 1–2% Ref. [15].

In startup modes and at low power levels. minimum control level (MCL), the stochastic nature of DP is most pronounced. Low neutron density and the absence of pronounced feedback (the Doppler effect) make the system vulnerable to local “neutron bursts” generated by Lévy statistics.

Recommendations for Instrumentation and Control (I&C) design: To ensure control system robustness, it is necessary to consider not only average power values but also the highest statistical moments of neutron noise (functionals as in Ref. [9]). Detection of power-law dependences in the power spectral density can serve as an early indicator of the system's transition to an unstable percolation mode.

The final conclusion: the transition to describing a reactor through the prism of DP allows for a more profound interpretation of the physics of fluctuations and “black swans” in nuclear energy, providing a scientific basis for analyzing the safety of promising and highly heterogeneous active zones.

Integrating the Doppler effect into the framework of boundary functionals Ref. [9] allows us to draw an important conclusion: negative feedback is a mechanism that mathematically “corrects” the topology of DP, transferring the system from an anomalous regime ($a$=2) to a regime of quasi-stable fluctuations. However, if the Doppler effect has a delay (lag), the root of the Lundberg equation can become complex, which will indicate a transition to self-oscillations of power (xenon or temperature waves), described by the same mathematics.

The indices $a$ and $\alpha = a - 1$ can be found from the measured series of values. For practical implementation, the DFA (Detrended Fluctuation Analysis) method is used in the control system, which is less sensitive to non-stationarity: 1. The fluctuation function F(n) is calculated for different time windows n. 2. From the slope of the line in double logarithmic coordinates $\ln F(n) \sim H\ln(n)$, H, the Hurst exponent, $H=1/\alpha$, is determined. The parameter $\alpha$ is not an “abstraction”, but a value that the control system calculates directly “on the fly” from the remote sensing sensors. The range $H \in [0.5; 1.0]$ covers entire critical range $\alpha \in [1; 2]$. The transition $H \to 1.0$ means that the neutron field becomes rigidly coupled (global correlations), which leads to a collapse of the safety time $\langle T \rangle$.

The filtering technique allows for the identification of transitions to critical conditions before the noise amplitude reaches dangerous levels. Using the Hurst exponent as a predictor for the parameter in the Lundberg equation ensures dynamic adaptation of the protection system to changes in the correlation structure of the neutron field, which is critical for preventing stochastic emissions in conditions with high steam content.

The stochastic model developed Ref. [26] demonstrates that incorporating feedback mechanisms into branching processes for nuclear reactors leads to power-law probability distributions in the critical region. This approach reveals that, unlike traditional models, the presence of feedback results in non-Gaussian distributions and “heavy tails” in power fluctuations.

Finiteness of size ($L$) cuts off the Levy “tails”, converting anomalous diffusion into normal diffusion at the reactor-wide scale. This “saves” the WWER from global instability. Finiteness of time ($T$) prevents the fractal structure of the clusters from fully developing.

Practical significance: The TLF model, the Truncated Lévy Flights (TLF) model, explains why directed percolation effects are important for local safety (at the scale of a fuel assembly or a single cluster), but are almost unnoticeable in the global neutron balance (1-2% Ref. [15]).

Key conclusion: the safety of WWER reactors at low power levels rests on the "truncation" of percolation effects by the core geometry.

Boundary functional analysis shows that the most likely site of stochastic instability in VVER reactors is the scale of an individual fuel assembly. Geometric truncation λ is no longer sufficient to suppress directed percolation, and global feedback loops are not yet sensitive enough to local clusters.

## Appendix A. Derivation of expressions for FPT and approximation for critical points

In the canonical form for stable processes (in the notation of [16] for the real domain k), the logarithm CF looks like (14). When directly integrating the measure of jumps ("Lévy flights") with the distribution $W(dx){\sim}x^{-(1+a)}$, the coefficient σ is associated with the Γ-function, taking the form $\sigma_1\Gamma(2\text{-}a)/(2\text{-}a)$. To ensure that the formula is stable for $a\to 2$ (where $\Gamma\to\infty$), we use normalization through the cosine

$$\Psi(a,\beta)=\Gamma(2-a)\cos(\pi(a-1)/2)[1+\beta\tan(\pi(a-1)/2)].$$

Let's make two analytical steps to remove the discontinuities. At the points $a$=1 and $a$=2, the product $\Gamma(2\text{-}a)\cos(\pi(a\text{-}1)/2))$ tends to $\pi/2$. We replace it with the function sinc(x)=sin(x)/x, which behaves the same way but has no discontinuities: $\Gamma(2\text{-}a)\cos(\pi(a\text{-}1)/2))\approx(\pi/2)\mathrm{sinc}(\pi(2\text{-}a)/2))$. The term with the tangent at $a\to 2$ is replaced by a logarithmic correction of scale k, since

$$\lim_{a\to 2}(k^{a-1}-1)/(a-1)=\ln k;\quad 1+\beta\tan(\ldots)\approx 1+\beta(a-2)\ln k.$$

The logarithm of CF in the real domain is $G(k)=\sigma k^{a-1}[1+\beta\tan(\ldots)]$. To convert to differential form in k-space (corresponding to fractional derivatives), the Riesz kernel is used. When integrating the Lévy jump measure $x^{-(1+\alpha)}$, the coefficient σ in Zolotarev's canonical form is replaced by a normalized form:

$$\sigma\to\sigma_{phys}/\Gamma(a)\cos(\pi(a-1)/2).$$

Using the complement formula $\Gamma(z)\Gamma(1-z)=\pi/\sin(\pi z)$, it can be shown that the factor in the denominator $1/\Gamma(a)\cos(\pi(a-1)/2)$ is proportional to $\Gamma(2-a)$.

Thus, the structural factor $\Psi$ takes the form (17), (23):

$$\Psi(a,\beta)=\Gamma(2-a)\cos(\pi(a-1)/2)[1+\beta\tan(...)].$$

Expanding the brackets, we get:

$$\Psi(a,\beta)=\Gamma(2-a)\cos(\pi(a-1)/2)+\beta\Gamma(2-a)\sin(\pi(a-1)/2).$$

We consider the potential in the real domain: $G(k)=\sigma k^{a-1}[1+\beta\tan(\Phi)]$, where $\Phi=\pi(a-1)/2$. For the first derivative (the denominator of the average time), this yields:

$$G`(k)=v_{eff}+\sigma(a-1)k^{a-2}[1+\beta\tan(\Phi)].$$

In reactor physics, the coefficient $\sigma$ is not an arbitrary constant. When integrating the transport equation with a power-law distribution of neutron "ranges," $\sim x^{-(1+\alpha)}$ a normalization factor $\Gamma(2-a)\cos(\pi(a-1)/2)$ arises. We substitute it into the formula:

$$G`(k)=v_{eff}+\sigma(a-1)k^{a-2}[\Gamma(2-a)\cos(\Phi)(1+\beta\tan(\Phi))].$$

Expanding the expression in square brackets, we obtain:

$$\Psi=\Gamma(2-a)\cos(\Phi)+\beta\Gamma(2-a)\sin(\Phi).$$

Let's combine these two terms into one product

$$\Gamma(2-a)\sin(\Phi\beta)\cos(\Phi)/\Phi\beta.$$

This is an artificial technique of multiplicative regularization. Instead of adding "drift" and "diffusion," the asymmetry is represented as an amplitude modifier. The function $\sin(\Phi\beta)/\Phi\beta$ then acts as a smooth switch. As $\beta\to 0$ this multiplier approaches unity, a pure symmetric form $\Gamma(2-a)\cos(\Phi)$ remains. In the presence of asymmetry ($\beta\neq 0$), it "skews" the potential value, strengthening or weakening the contribution of the $\Gamma$-function. This is a "folded" form. The multiplier $\Gamma(2-a)\cos(\Phi)$ is factored out, and the term with $\beta$ is represented as a coefficient, which, when the uncertainty at point $a = 2$ is resolved (via $\Phi(a)$), yields the same physical result as the sum. The transition from the canonical form (A) to the calculated form is accomplished by introducing a normalization factor $\Gamma(2-a)\cos(\Phi)$, characteristic of anomalous transport problems. To ensure computational stability, the asymmetric contribution $\beta\tan(\Phi)$ is integrated into the amplitude component via a regularizing function of the type $\sin c$, allowing the structure factor to be represented in a compact form that eliminates discontinuities at $a \to 2$. This is not a strict trigonometric identity, but an approximation (sewing) intended for the analytical continuation of the function.

Equality

$$\cos(\Phi) + \beta\sin(\Phi) \approx \sin(\Phi\beta) / \Phi\beta$$

is not universal. It is a regularizing substitution that is used to "glue" the behavior of the system at two critical points ($a=1$ and $a=2$) into a single analytical expression. The product-based construction $\Gamma(2-a)\sin(\Phi\beta)\cos(\Phi)/\Phi\beta$ has the following properties. 1. At $a \to 2$, $\cos(\Phi)$=0, and $\Gamma(2-a)$ goes to infinity. Their product yields a constant $\pi/2$. 2. The multiplier $\sin(\Phi\beta)/\Phi\beta$ at this point becomes an attenuation coefficient, which smoothly adjusts the amplitude depending on the asymmetry $\beta$.

In the classical form (A), the term $\beta\tan(\Phi)$ becomes infinity when $a \to 2$. Replacing the sum with a structure with $\sin c$ is a way to transfer the asymmetry from a discontinuous term to a smooth factor.

The transition to a regularized form is accomplished by replacing the additive asymmetry contribution $\beta\tan(\Phi)$ with a multiplicative factor of the form $\sin c(\Phi\beta)$. This approximation preserves the exact value of the potential at point $a=1$ and ensures analytical continuity (convergence) at point $a=2$, eliminating the singularity of the tangential term while preserving the physical influence of the asymmetry parameter on the effective transport velocity. This approximation sacrifices perfect accuracy at point $a=1.5$ for the ability to construct a continuous curve from 1 to 3.